\documentclass[twocolumn,twoside]{revtex4}
\usepackage{graphicx}
\usepackage{fancyhdr}
\begin{document}

\title{Lepton flavour violation in muon decays}

%

\author{L.~Galli}
\affiliation{INFN Sezione di Pisa, Largo B. Pontecorvo 3, 56127, Pisa}

\begin{abstract}
The search for lepton flavour violation in charged lepton decays is highly sensitive to physics beyond the Standard Model. Among the possible processes, $\mu$-decays are considered to have the largest discovery potential in most of the standard model extensions.  Many searches has been performed in the past, however no evidence has been found so far. Four dedicated experiments are in advanced state of preparation to improve the current associated sensibilities by 1-4 order of magnitudes for the charged lepton flavour violating processes $\mu\rightarrow \rm{e}\gamma$, $\mu\rightarrow \rm{e}$ conversion and $\mu\rightarrow \rm{eee}$. In this paper I present  physics motivations, experimental challenges and  construction status of the experiments, which are the studying above mentioned processes.
\end{abstract}

\maketitle

\thispagestyle{fancy}


\section{Introduction}

This proceeding is a short review about charged lepton flavour violation in the muon sector. The reader interested in more details is invited to look at \cite{sig,cei,ootani,bernstein}.

In the minimal Standard Model (SM) the charged lepton flavour violating processes (cLFV) are prohibited since the doublets are separated and the neutrinos are massless. In this framework the three lepton families are separated and the lepton flavour number is separately conserved in any processes. Despite the fact that the neutrino oscillation phenomena show that neutrino are definitely massive , and even taking into account this effect, the branching fraction associated to a cLFV decay  $\mu\rightarrow\rm{e}\gamma$ in the SM is unmeasurably small $\approx 10^{-54}$:

\begin{equation}
\mathcal{B}(\mu\rightarrow\rm{e}\gamma) = \frac{3\alpha}{32\pi}\large{|} \sum_{i=2,3} U^*_{\mu i} U_{ei} \frac{\Delta m^2_{i1}}{M_W^2} \large{|}^2
\label{eq:SM}
\end{equation}

\begin{figure}[h]
\centering
\includegraphics[width=80mm]{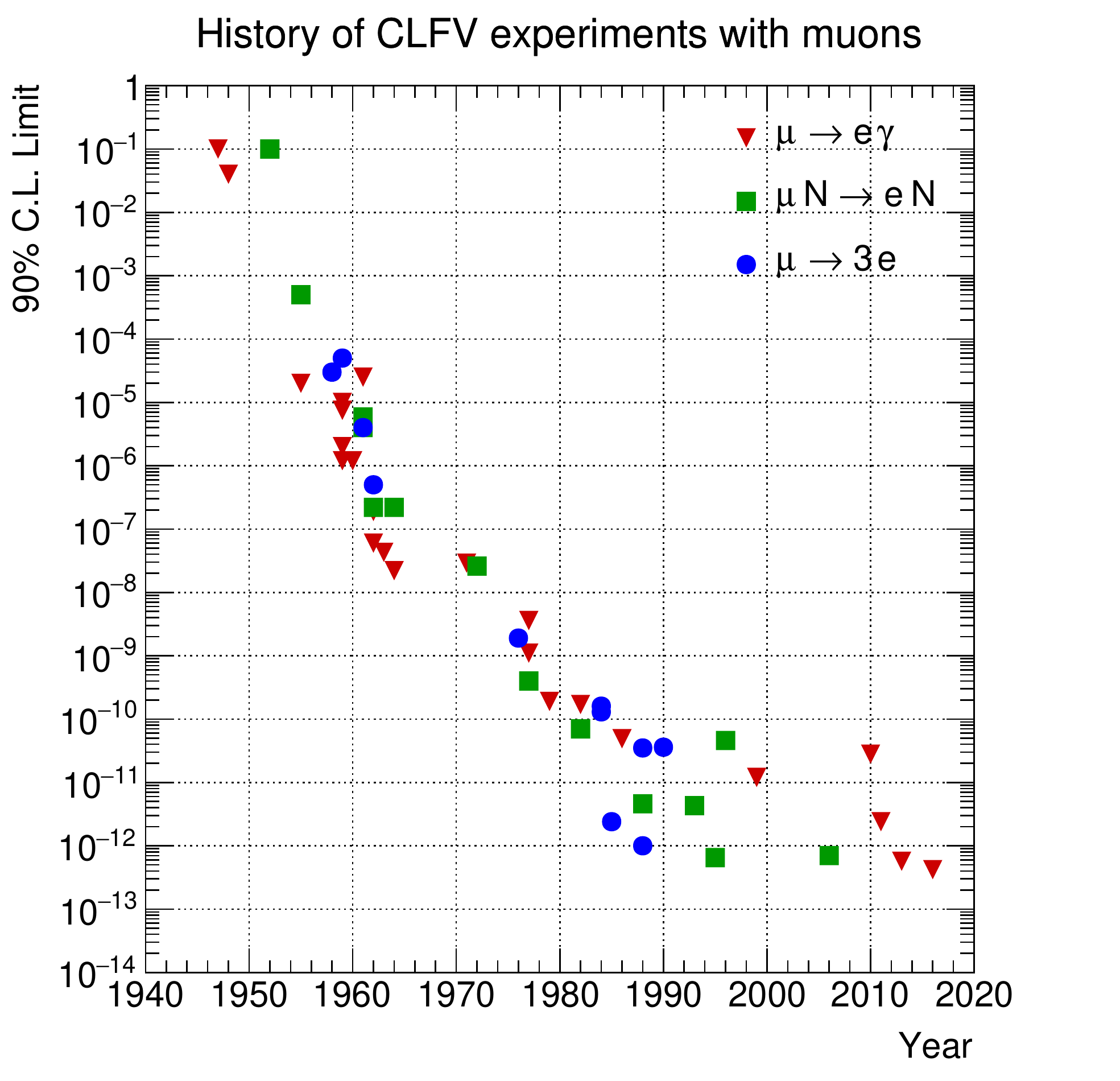}
\caption{cLFV upper limits history in the last 80~years.} \label{fig:clfv}
\end{figure}

Thus any evidence of $\mu\rightarrow\rm{e}\gamma$ decay would incontrovertibly demonstrate the existence of physics beyond the Standard model (BSM).  Figure~\ref{fig:clfv} shows the evolution of the sensitivities on the cLFV processes that has been reached in the last 80~years. The lack of the signal was one of the cornerstone of the leptonic flavour structure in the SM. See~\cite{kuno} as a general reference.

Muons are very sensitive probes of cLFV, in fact intense muon beams can be obtained by hitting light targets with low energy protons ($\le$~590~MeV/c at Paul Scherrer Institut, PSI) or at proton accelerators as by product of high energy collisions (Fermilab and J-PARC); the relatively long decay time, 2.2 $\mu$sec, allows to transport those beams on a thin target to let the muons decay at rest. It is possible to accumulate large amount of statistics and then reach sensitivities in the range 10$^{-14}\div 10^{-17}$ as it will be described in details in the following sections together with a comparison among the various decays and their discovery power.

\subsection{cLFV and physics beyond the standard model: a model independent approach}

The SM is widely considered a low energy approximation of a more general theory, possible extensions are associated to theories, among many others, such as super-simmetry, grand unification of the forces and the Majorana nature of the neutrino. Any of the before mentioned frameworks produces a prediction of cLFV in the range accessible to experiments as a function of the theory parameters: couplings and energy scale. 

\begin{figure}[t]
\centering
\includegraphics[width=80mm]{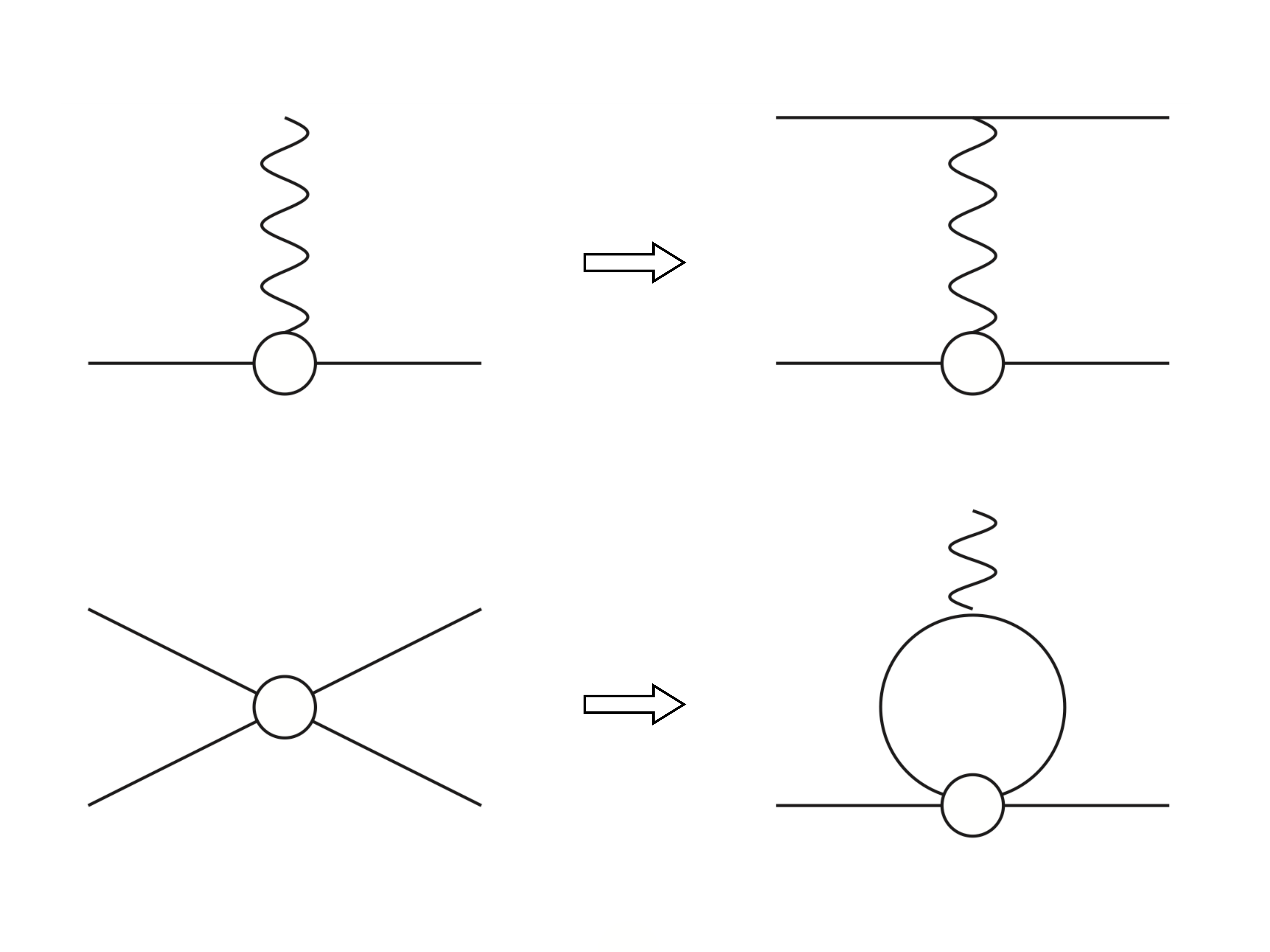}
\caption{Schematic representation of the tree level and associated one-loop contribution for $\mu\rightarrow\rm{e}\gamma$ decay in the top row and $\mu\rightarrow \rm{eee}$ and $\mu\rightarrow\rm{e}$ conversion in the bottom. The BSM interaction is in the circle.} \label{fig:fey}
\end{figure}

Independently of the nature of the BSM physics cLFV in the $\mu$-decay would happen via a dipole transition or a phenomenological 4-fermion like interaction as schematically represented in Figure~\ref{fig:fey}. 

The diagrams schematically show the dipole operator contributes at one-loop order to $\mu\rightarrow \rm{eee}$ and $\mu\rightarrow\rm{e}$ and vice versa how the 4-fermion works for $\mu\rightarrow\rm{e}\gamma$; thus a combined evidence of cLFV from more than one experiment can indicate the nature of the SM extension; the experiments presented below are in this sense complementary. 

\begin{figure}[t]
\centering
\includegraphics[width=80mm]{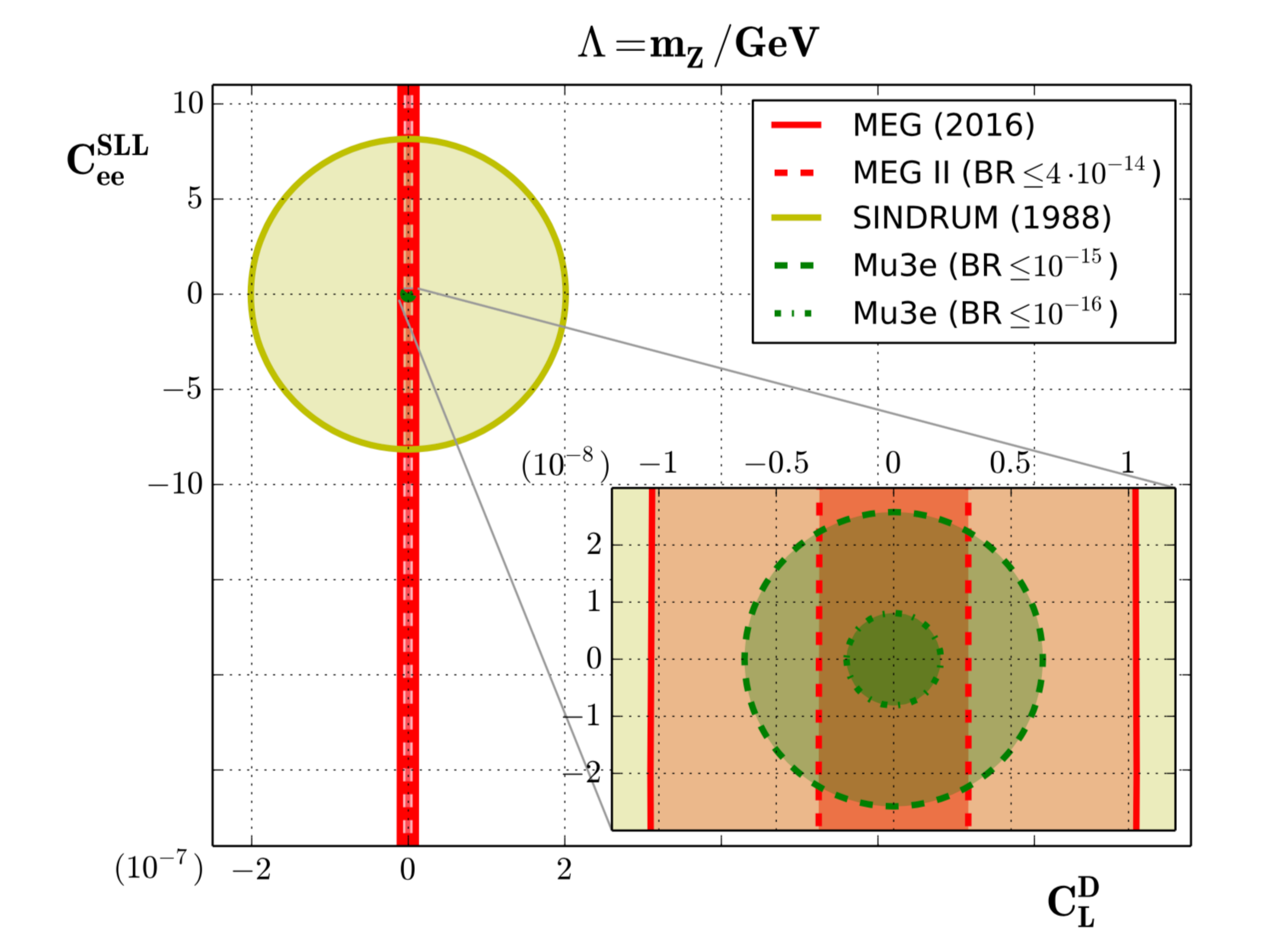}
\caption{Allowed values for the dipole and 4-fermion couplings given by current and future limits on $\mu\rightarrow\rm{e}\gamma$ and $\mu\rightarrow \rm{eee}$ decays~\cite{pruna}.} \label{fig:comp}
\end{figure}

An example is shown in Figure~\ref{fig:comp} where the allowed values for dipole couplings and for 4-fermion interaction are shown given present and future limits on the $\mu\rightarrow\rm{e}\gamma$ and $\mu\rightarrow \rm{eee}$ decays~\cite{pruna}.

\section{Experimental searches: generalities}

Given the energy-momentum conservation only three processes are considered for the cLFV, each one deserves a dedicated experiment with peculiar characteristics. All searches differ in terms of beam intensity, structure, detection technique and related backgrounds. 

In general, in order to reach the experimental sensitivity, the experiments are equipped with detectors having resolution at unprecedented values for particles (e and $\gamma$) of energy in the range 10-100~MeV, with fast response to cope with $\mu$-decay intensity and as transparent as possible to minimise any matter effect. Ideally, if one was able to measure the experimental observables and to eliminate any beam induced background, these searches would be background-free searches; this is the cLFV experimental challenge. 

\subsection{The $\mu\rightarrow\rm{e}\gamma$ search in MEG~II}

MEG~II at Paul Scherrer Institute searches for the $\mu^+\rightarrow\rm{e^+}\gamma$ decay with a design sensitivity of 6$\times10^{-14}$\cite{MEGII}, about one order of magnitude lower with respect to the previous result obtained by the same collaboration with the MEG experiment\cite{MEG}. Positive muons are used to avoid interferences due to the negative muon capture in nuclei.

The signature is a time coincident, back-to-back pair of a monoenergetic photon and a monoenergetic positron, both with an energy equal to half of the muon mass, $\rm{E_{e}=E_{\gamma}}\approx$~52.8~MeV.  There are two major backgrounds: one is a prompt background from radiative muon decay, $\mu^+\rightarrow\rm{e^+ \nu_e \overline{\nu}_{\mu}}\gamma$, when $\rm{e^+}$ and the $\gamma$ are emitted back-to-back with the two neutrinos carrying away little energy; in this decay the two particles are emitted at the same time. The other background is an accidental coincidence of a $\rm{e^+}$ in a normal $\mu$-decay, $\mu^+\rightarrow\rm{e^+ \nu_e \overline{\nu}_{\mu}}$, accompanied by a high-energy photon from a muon radiative decay or a positron annihilation in flight. The accidental background gives the major contribution since the associated rate increases with the square of the muon beam intensity; the signal to noise ratio is then maximised by using a continuous beam. 

\begin{figure}[t]
\centering
\includegraphics[width=80mm]{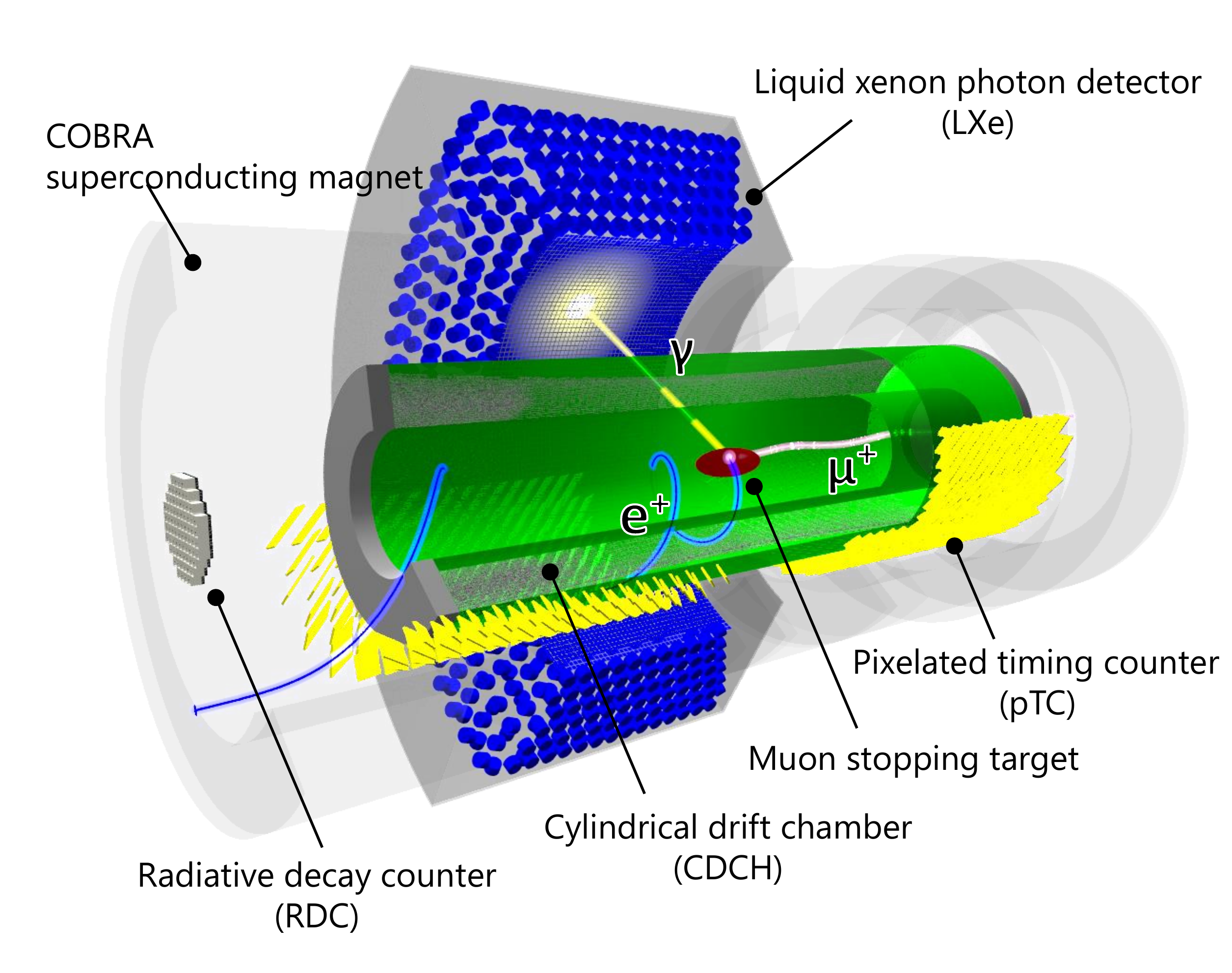}
\caption{Schematic view of the MEG~II detector.} \label{fig:meg2}
\end{figure}

A $7 \times 10^{7}$ $\mu^{+}$/s beam is stopped in a $140~{\rm \mu m}$ slanted polyethylene target. The ${\rm e}^{+}$ momentum is measured by a magnetic spectrometer, composed by an almost solenoidal magnet (COBRA) with an axial gradient field and by an ultra thin unique volume cylindrical drift chamber. The ${\rm e}^{+}$ timing is measured by two matrices of 256 plastic scintillators pixels read out with SiPMs (Timing Counter, TC). The $\gamma$ energy, direction and timing are measured in a $\approx 800~{\rm l}$ volume liquid xenon (LXe) scintillation detector by means of MPPCs on the inner face and PMTs on the lateral. A schematic overview of the MEG~II detector is shown in Figure~\ref{fig:meg2}.

The experiment is in an advanced construction phase. The detector integration has been completely tested in 2018, and at the end of 2019 an engineering run is scheduled; the beginning of DAQ collection is foreseen in 2020. Three full years of data taking are expected to reach the experimental goal within 2023.

\subsection{The $\mu\rightarrow\rm{eee}$ deacy in Mu3e}

The signal is the emission of two positrons and an electron on a plane, from a common vertex with a total momentum equal to $\vec{0}$ and a total energy equal to the muon mass $\rm{E_{tot}}\approx$~105.6~MeV. Being a three-body decay the energy of the daughters is not a fixed values, being the higher energy larger than 35~MeV while the lowest energy peaks near zero and about one half have an energy larger than 15~MeV. The detector must have an excellent tracker as thin as possible in order to have high acceptance for tracks from few MeVs to half of the muon mass. Similarly to the  $\mu\rightarrow\rm{e}\gamma$ a positive muon beam is chosen and there are two sources of background. The prompt is given by the $\mu\rightarrow\rm{e^+e^-e^+\nu_e\overline{\nu}_{\mu}}$ where the two neutrinos carry very little energy. The other background comes from the accidental coincidence of two or three muon-decays, and strongly depends on the beam rate, for this reason a continuous beam will be used.

The Mu3e experiment~\cite{mu3e} in under construction at Paul Scherrer Institut and aims at reaching a 10$^{-16}$ sensitivity in two successive phases and improving the former result by 3 orders of magnitudes~\cite{SINDRUM}. The muon beam will be transported to a double cone Mylar target, 85$\mu$m thick. The target is placed in a solenoidal field of 1~T, the detector is made of ultra-thin (50~$\mu$m) silicon pixels and scintillators fibers and tiles read out by means of SiPMs. 

In order to reach the experimental goal a new beam-line concept is required: a dedicated R\&D called HiMB is presently ongoing at PSI in order to have muon beam intensities up to few $10^9$~$\mu$/sec~\cite{himb}.

The detector commissioning for the Phase-I is foreseen for 2021, the physics program will last until 2030 with the Phase-II.

\begin{figure}[t]
\centering
\includegraphics[width=80mm]{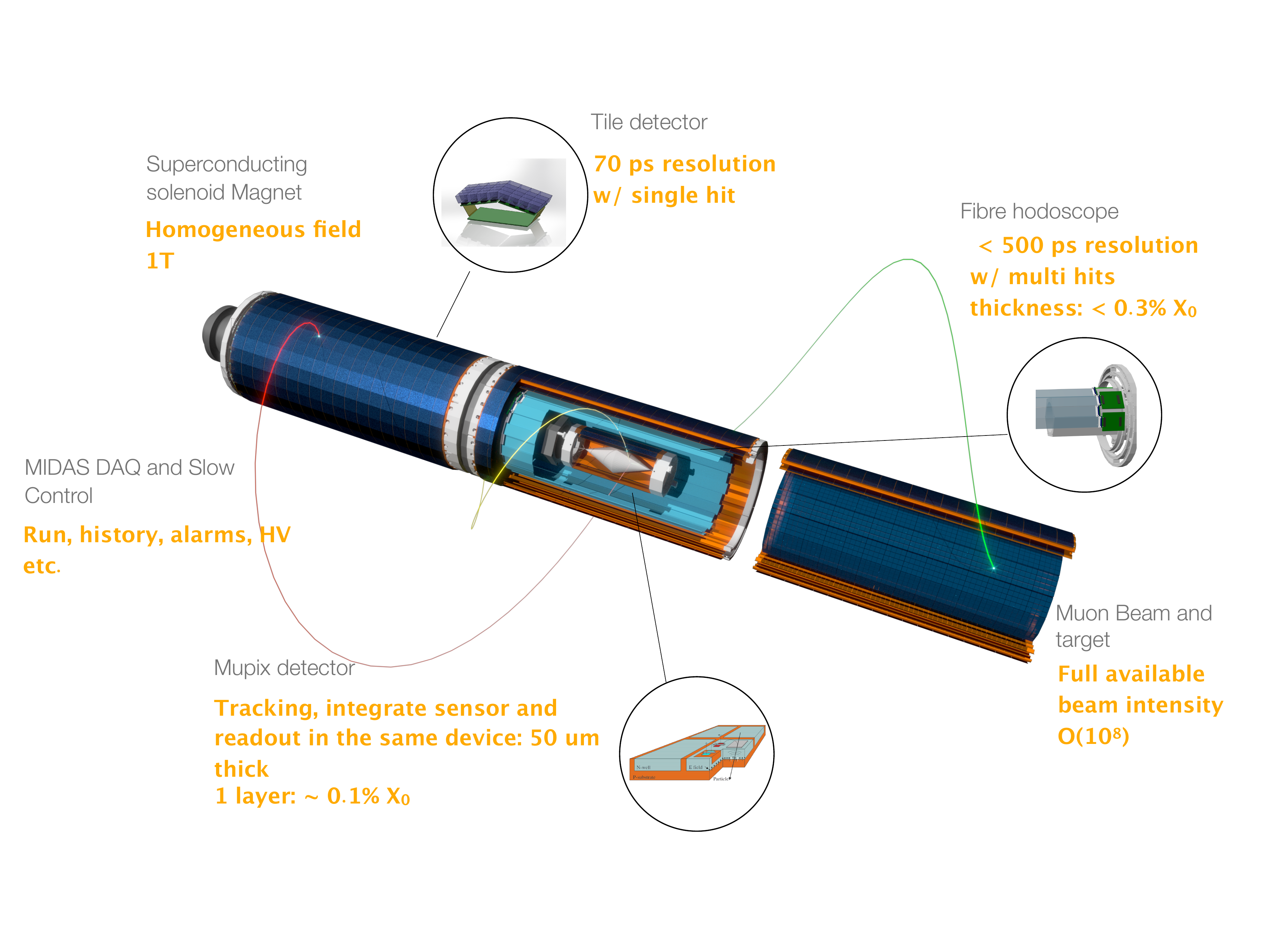}
\caption{Schematic view of the Mu3e detector.} \label{fig:mu3e} 
\end{figure}

\subsection{The $\mu\rightarrow\rm{e}$ conversion}

The $\mu\rightarrow\rm{e}$ conversion is a cLFV process which can take place when negative muons are stopped in the matter through the process $\mu^- + \rm{(A)} \rightarrow \rm{e^- (A)}$, where A is a nucleus like for example Al. A pulsed negative muon beam is formed from the decay of pions produced in proton collision on fixed target and brought to stop in a layer of thin targets, where muon capture can take place. The signal is a final state with one electron having the energy equal to the muon mass reduced by the difference in binding energy of the nucleus before and after the reaction. Since this is a single particle final state process the sensitivity is not limited by the accidental background which does not pose in principle any limitation in the muon intensity. A beam induced background can originate from the interaction of pions in the beam lines with the target; this can be eliminated by implementing a highly asymmetric proton beam structure with very short and intense proton spills well separated in time. All the pions created in the proton collisions decay within few hundred ns, after that  electrons can be emitted only by muon-related decays, the distance between the proton spill is of order of 1/2~$\mu$s. The residual fractional contribution of out of time protons (the so called extinction factor), and then of pions, has to be at least of $10^{-9}$.  Other backgrounds come from the interaction of cosmic rays in the target and the muon decay in flight. 

The current sensitivity is a few 10$^{-13}$ depending on the target used and was obtained by the SINDRUM-II experiment at Paul Scherrer Institut~\cite{SINDRUMII,SINDRUMII2}. Two new experiments are currently under construction aiming at improving the sensitivity by 4 orders of magnitudes.

\subsubsection{The $\mu\rightarrow\rm{e}$ in Mu2e}

\begin{figure}[t]
\centering
\includegraphics[width=80mm]{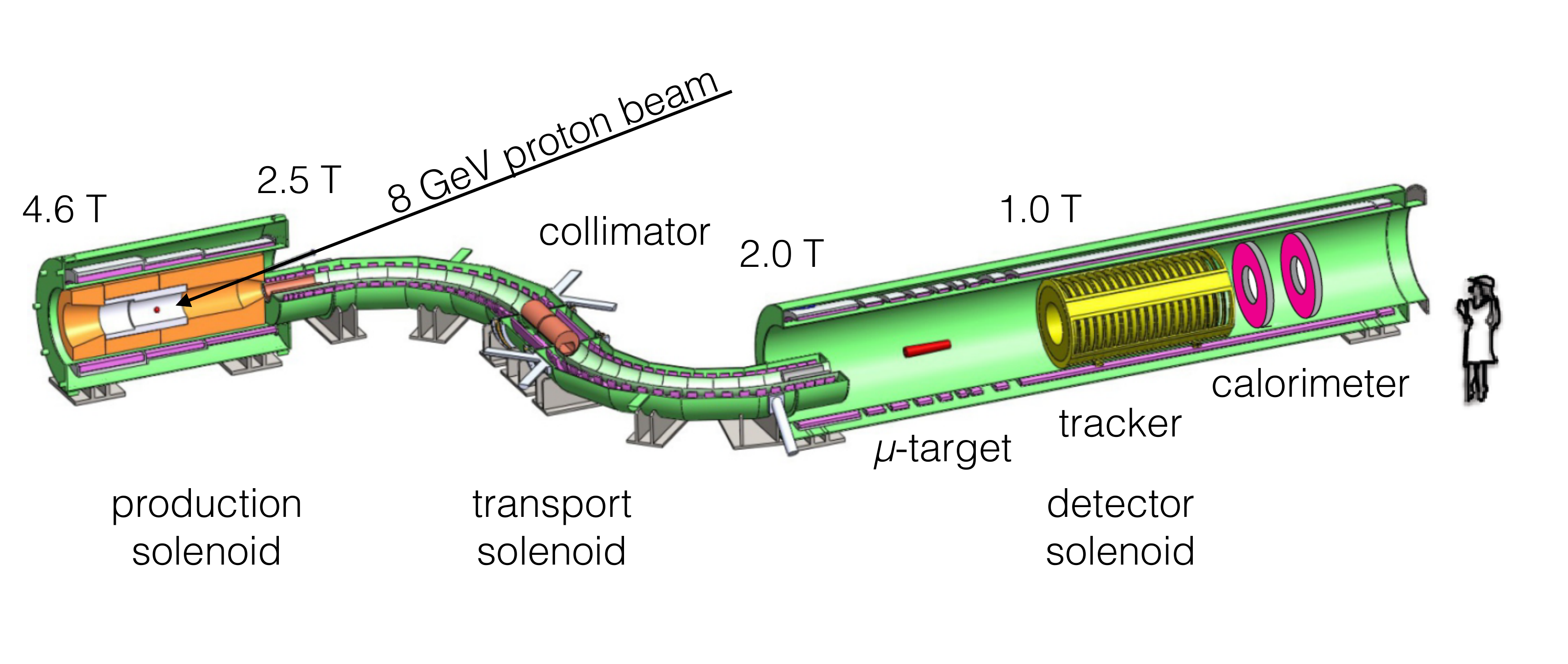}
\caption{Schematic view of the Mu2e beam line and detector.} \label{fig:mu2e}
\end{figure}

The Mu2e experiment is currently under construction at the muon campus in Fermilab National Laboratory~\cite{mu2e}. The backward-going pions produced by the 8~GeV proton beam are captured and decay into muons, which are transported through a bent solenoid to a series of aluminum disks where they stop. The electron coming from the muon decay or capture is measured by a straw tube tracker and a pair of crystal calorimeters, arranged in the shape of a hollow cylinder to let low momentum electrons go through undetected. Figure~\ref{fig:mu2e} shows a schematic view of the beam line and detector.

Mu2e is in advanced state of commissioning, it is expected to start data taking in 2023 and to reach 90\% confidence level sensitivity of $6\times10^{-17}$ in three years of data taking.

\subsubsection{The $\mu\rightarrow\rm{e}$ in COMET}

\begin{figure}[t]
\centering
\includegraphics[width=80mm]{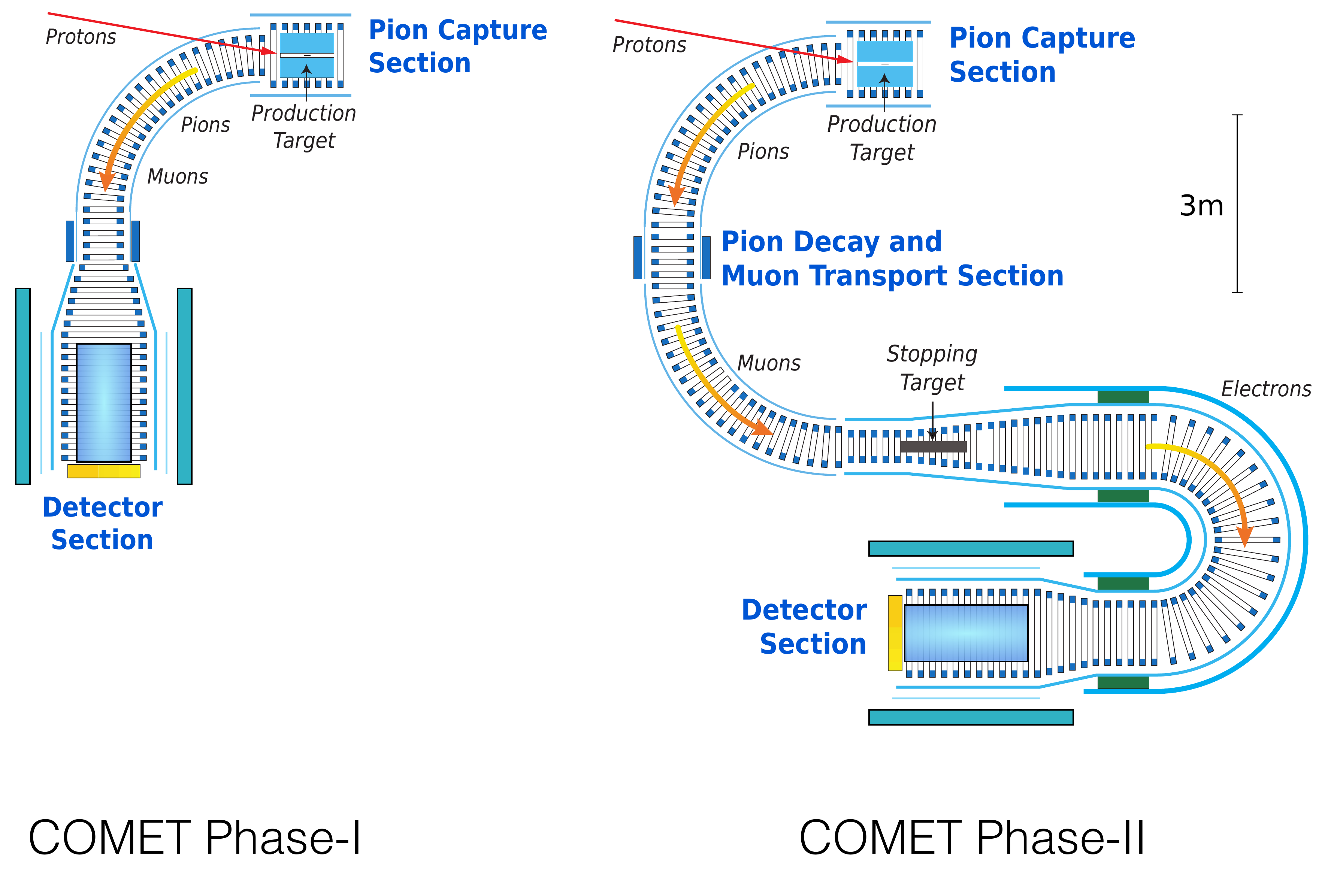}
\caption{Schematic view of the COMET Phase I and II beam lines and detectors.} \label{fig:comet}
\end{figure}

The COMET (COherent Muon-to-Electron Transition) experiment is in construction at the Japanese Proton Accelerator Research Center (J-PARC)~\cite{comet}. COMET will operate in two stages, Phase-I and Phase-II that are schematically reported in Figure~\ref{fig:comet}. Phase-I has a sensitivity goal of $2\times10^{-15}$ and will help to understand some of the novel experimental techniques, the beam and the background rates. In the first experimental stage the target will be placed at the center of a thin cylindrical drift chamber surrounded by scintillating hodoscopes for triggering and timing. In the Phase-II the beam line will be extended and the electrons will be tracked by a forward straw tube tracker and a calorimeter made of LYSO crystals. Phase-II will allow to reach a sensitivity of $3\times10^{-17}$. Recent measurements have been able to demonstrate an extinction factor rate of about 10$^{-12}$.
\\

\vspace{20cm}

COMET Phase-I is currently under construction together as the facility. The physics data taking is expected in 2021/2022 and will take 2/3 years to accumulate the statistics. Phase-II will follow.

\section{Conclusions and prospects}

\begin{figure}[t]
\centering
\includegraphics[width=80mm]{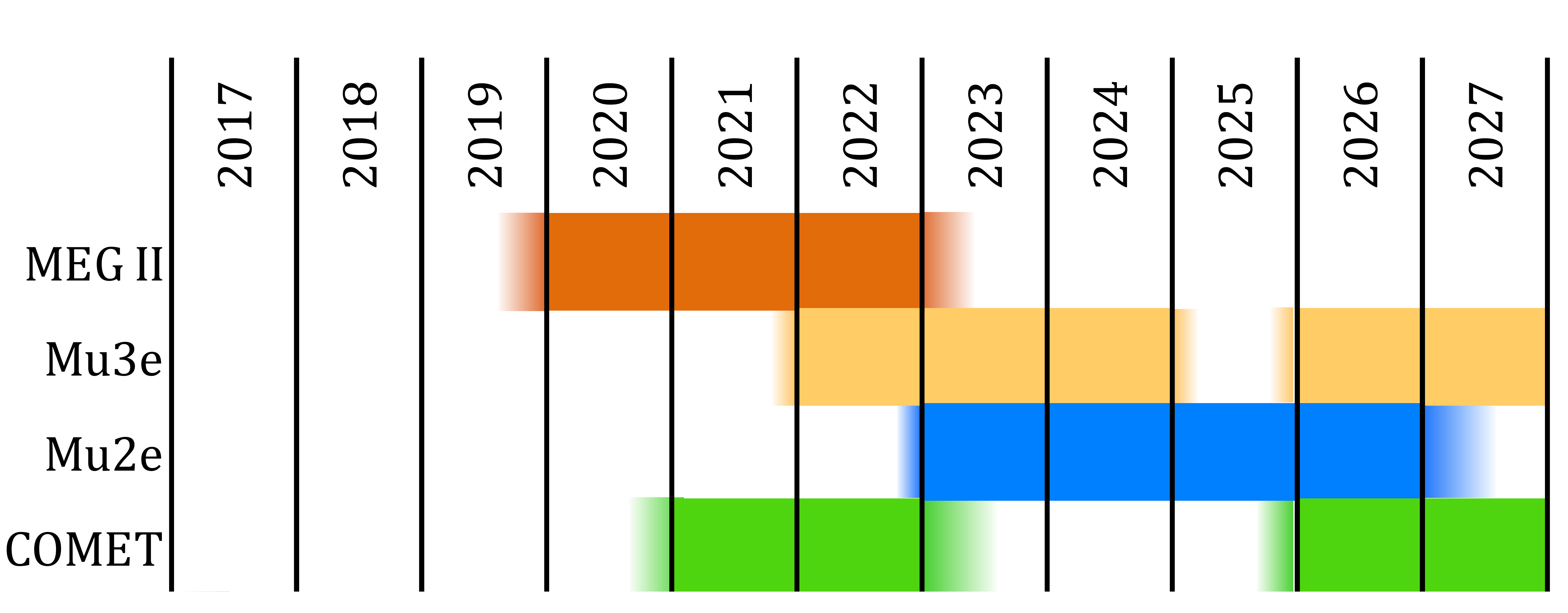}
\caption{Current schedule of the search of cLFV in the muon sector.} \label{fig:timeline}
\end{figure}

It is exciting to see that within the next five to ten years our present knowledge of fundamental interactions could be disproved or confirmed by means of a full set of complimentary searches of BSM physics in the muon sector. As shown in Figure~\ref{fig:timeline} MEG~II is expected to start data taking in 2020 after an engineering run in 2019; Mu3e Phase-I commissioning is expected for 2021 followed by almost 10 years of physics program towards the Phase-II; Mu2e foresees three years of data taking starting in 2023; finally COMET Phase-I data collection is expected to start in 2021/2022.

Table~\ref{tab:sens} reports the current cLFV sensitivities in the muon sector and how it will be improved by the new generation of experiments described.

\begin{table}[h]
\begin{center}
\caption{Present and future sensitivities in the cLFV in the muon sector.}
\begin{tabular}{|l|c|c|}
\hline \textbf{Process} & \textbf{Present} & \textbf{Future} \\
\hline $\mu\rightarrow\rm{e}\gamma$ & 4.2$\times10^{-13}$ & 6$\times10^{-14}$  \\
\hline $\mu\rightarrow\rm{e}$ conversion  & $10^{-12}\div10^{-13}$ & $10^{-17}$ \\
\hline $\mu\rightarrow\rm{eee}$   & $10^{-12}$ & $10^{-16}$\\
\hline
\end{tabular}
\label{tab:sens}
\end{center}
\end{table}

\end{document}